\documentclass[]{aa}
\usepackage{graphicx}
\usepackage{color}
\newcommand{\beq}{\begin{equation}}
\newcommand{\eeq}{\end{equation}}
\newcommand{\beqa}{\begin{eqnarray}}
\newcommand{\eeqa}{\end{eqnarray}}
\begin{document}
\title{Small-scale H$\alpha$ Jets in the Solar Chromosphere}

\author{D.Kuridze$^{1}$, M.Mathioudakis$^{1}$, 
D.B.Jess${^1}$, S.Shelyag$^{1}$, D.J.Christian$^{2}$, F.P.Keenan${^1}$, K.S.Balasubramaniam$^3$}
\institute{Astrophysics Research Centre, School of Mathematics and 
Physics, Queen's University Belfast, Belfast, BT7 1NN, Northern Ireland, U.K. 
\and Department of Physics and Astronomy, California State University, 
Northridge, CA 91330, USA
\and Air Force Research Laboratory, Solar and Solar Disturbances, Sunspot, NM 88349, USA}

\date{received / accepted }

\abstract {}
{High temporal and spatial resolution observations from the Rapid Oscillations in the 
Solar Atmosphere (ROSA) multiwavelength imager on the Dunn Solar Telescope are used to study the velocities of small-scale H$\alpha$ jets in an emerging solar active region.}
{The dataset comprises of simultaneous imaging in the H$\alpha$ core, Ca II K, and G band, together with photospheric 
line-of-sight magnetograms. Time-distance techniques are employed to determine projected plane-of-sky velocities.}
{The H$\alpha$ images are highly dynamic in nature, with estimated jet velocities as high as 45~km s$^{-1}$. 
These jets are one-directional, with their origin seemingly linked to underlying Ca II K brightenings and G-band magnetic bright points.}
{It is suggested that the siphon flow model of cool coronal loops is suitable for the interpretation of our observations. 
The jets are associated with small-scale explosive events, and may provide a mass outflow from the photosphere to the corona.}

\titlerunning{Small-scale H$\alpha$ Jets in the Solar Chromosphere}
\authorrunning {Kuridze et al.}
\maketitle

\section{Introduction}
The solar chromosphere is permeated by a wide range of highly dynamic features 
including jets, fibrils, mottles, spicules, Ellerman bombs and 
H$\alpha$ surges. In particular, chromospheric jets are one of the most important, 
yet also most poorly understood phenomena of the Sun's magnetic atmosphere. 
Observed chromospheric jet velocities are generally measured to have values  
of around ${\sim}20$--$40$ $\mathrm{km\, s}$$^{-1}$  
(Tsiropoula \& Tziotziou \cite{tsiropoula}; Chae et~al. \cite{chae}; Lin et~al. \cite{lin}). 
Apart from these typical speeds, high velocity flows ($>$$100$ $\mathrm{km\, s}$$^{-1}$) are also observed in some 
chromospheric surges and blobs (Foukal \cite{foukal}; van Noort et~al. \cite{noort}). 

Several previous studies indicate that the energy required to drive these chromospheric jets can be 
released by magnetic reconnection occurring low in the 
solar atmosphere (Canfield et~al. \cite{canfield}; Chae et~al. \cite{chae1}; Chae \cite{chae2}; 
Shibata et~al. \cite{shibata}).
Yoshimura et~al. (\cite{yoshimura}) found a close correlation between H$\alpha$ surges, 
Transition Region and Coronal Explorer (TRACE) brightenings in the 1600 {\AA} waveband, and 
magnetic flux cancellations surrounding ephemeral regions. 
More recently, Morita et~al. (\cite{morita}) observed chromospheric 
anemone jets in Ca {\sc{ii}} K$_{1}$ and K$_{2}$ lines as intensity 
peaks in the light curves. They also found cancellation of magnetic flux near the jet area. 
These results strongly suggest that chromospheric 
jets may be generated by magnetic reconnection, similar to that occurring on a larger 
scale in the corona. Also, the 
multi-wavelength studies of Yoshimura et~al. (\cite{yoshimura}) and Brooks et~al. (\cite{brooks}) 
indicate that chromospheric 
surges are indeed driven by magnetic reconnection. Alternatively, 
Hansteen et~al. (\cite{hansteen}) suggest that the formation of jets, linked to dynamic 
fibrils, mottles and spicules, could be driven by magneto-acoustic shocks that leak 
upward through the chromosphere as a result of convective flows in the 
photosphere and global p-mode oscillations. 

Recently, a multi-wavelength study involving Solar Dynamic Observatory (SDO) and Hinode observations has 
highlighted the importance of chromospheric jets 
and spicules for providing the mass supply for the corona 
(De Pontieu et~al. \cite{depontieu}). However, spectroscopic observations 
can only provide the line-of-sight component of velocity, 
with any small-scale Doppler phenomena cancelling out if the spatial resolution 
is not sufficiently high. To investigate the small-scale structure, here  
we present high spatial and temporal resolution observations of an 
emerging active region in H$\alpha$ core, Ca {\sc{ii}} K, 
G band and line-of-sight magnetograms. 
In the following sections, we describe the details of the observations and the 
image processing (\S{2}), the analysis and results (\S{3}), 
and finally discuss and summarise possible interpretations of our findings (\S{4}).

\begin{figure*}[]
%\begin{center}
%\includegraphics[width=8.0cm]{m=0,n=2}
%\includegraphics[width=8.5cm]{m=0,n=1}
\includegraphics[width=17.5cm]{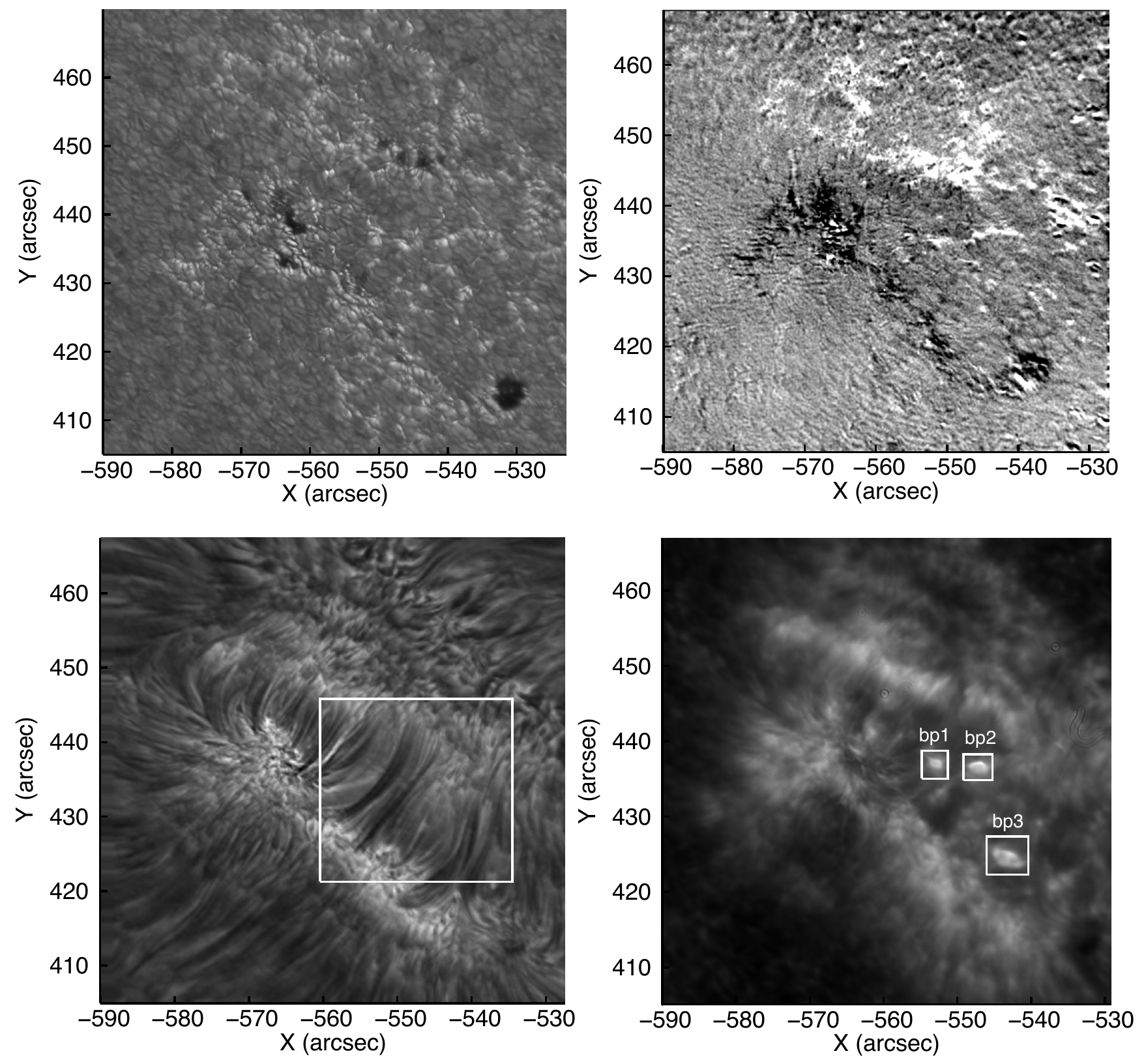}
%\includegraphics[width=9.0cm]{fig3}
%\end{center}
\caption{Simultaneous ROSA images of the G band (top left), line-of-sight magnetogram 
(top right),  H$\alpha$ core (bottom left), and Ca {\sc{ii}} K core (bottom right). 
The magnetogram colour scale is in Gauss, where white indicates positive magnetic 
polarities, while black demonstrates polarities which are negative. Artificial 
saturation at $\pm400$~G is implemented to highlight regions of weak magnetic polarity 
on the solar surface. The white rectangle in the H$\alpha$ core image indicates the 
approximate position where most of the jets are observed. White rectangles in the 
Ca {\sc{ii}} K image, identified by `bp1', `bp2', and `bp3', are the bright points 
selected for temporal analysis. Axes are in heliocentric 
arcseconds, where $1'' \approx 725$~km.}
\label{fig1}
\end{figure*}

\section{Observations \& Data Reduction}
The observations were obtained between approximately 15:00--16:00~UT on 2009 
May 31 with the Rapid Oscillations in the Solar Atmosphere (ROSA; Jess et~al. \cite{jess}) 
imaging system, mounted on the Dunn Solar Telescope (DST) at the National Solar 
Observatory, New Mexico, USA. High-order adaptive optics were used throughout 
the observations. The photospheric target under investigation comprises of a multitude 
of magnetic pores and associated faculae, located at heliocentric coordinates 
($-540''$, $435''$). Our dataset includes simultaneous imaging in the G band, H$\alpha$ core, 
Ca {\sc{ii}} K and line-of-sight (LOS) magnetograms. A spatial sampling 
of $0.069''$/pixel was selected, matching the diffraction limit of the DST to that of the 
G band, resulting in a field-of-view of approximately $69'' {\times} 69''$.

The images obtained were processed with speckle reconstruction algorithms 
(W\"{o}ger et~al. \cite{wger}), with the removal of large-scale seeing distortions achieved by destretching the data relative to simultaneous 
high-contrast continuum images. The algorithms 
utilised 32 images per reconstruction, resulting in an effective cadence of 8.4~s for 
H$\alpha$, Ca {\sc{ii}} K, and LOS magnetograms, while the reconstructed 
G-band cadence was 1.1~s. Magnetograms were 
constructed as normalised to their sum difference images of 
left- and right-hand circularly polarized light obtained 
125~m{\AA} into the blue wing of the magnetically-sensitive Fe~{\sc{i}} absorption line 
at 6302.5~\AA. A blue-wing offset was required to minimise granulation contrast, 
while conversion of the filtergram into units of Gauss was performed using simultaneous 
SoHO/MDI magnetograms (see discussion in Jess et~al. \cite{Jess10a}).  
ROSA images in G band, H$\alpha$ core, Ca {\sc{ii}} K and LOS 
magnetograms are shown in Fig.~\ref{fig1}. 
%The resulting images 
%were remapped to correct for foreshortening effects and to allow 
%for more accurate estimates of horizontal velocities.

\begin{figure*}[]
\vspace*{1 mm}
\begin{center}
\includegraphics[width=18 cm]{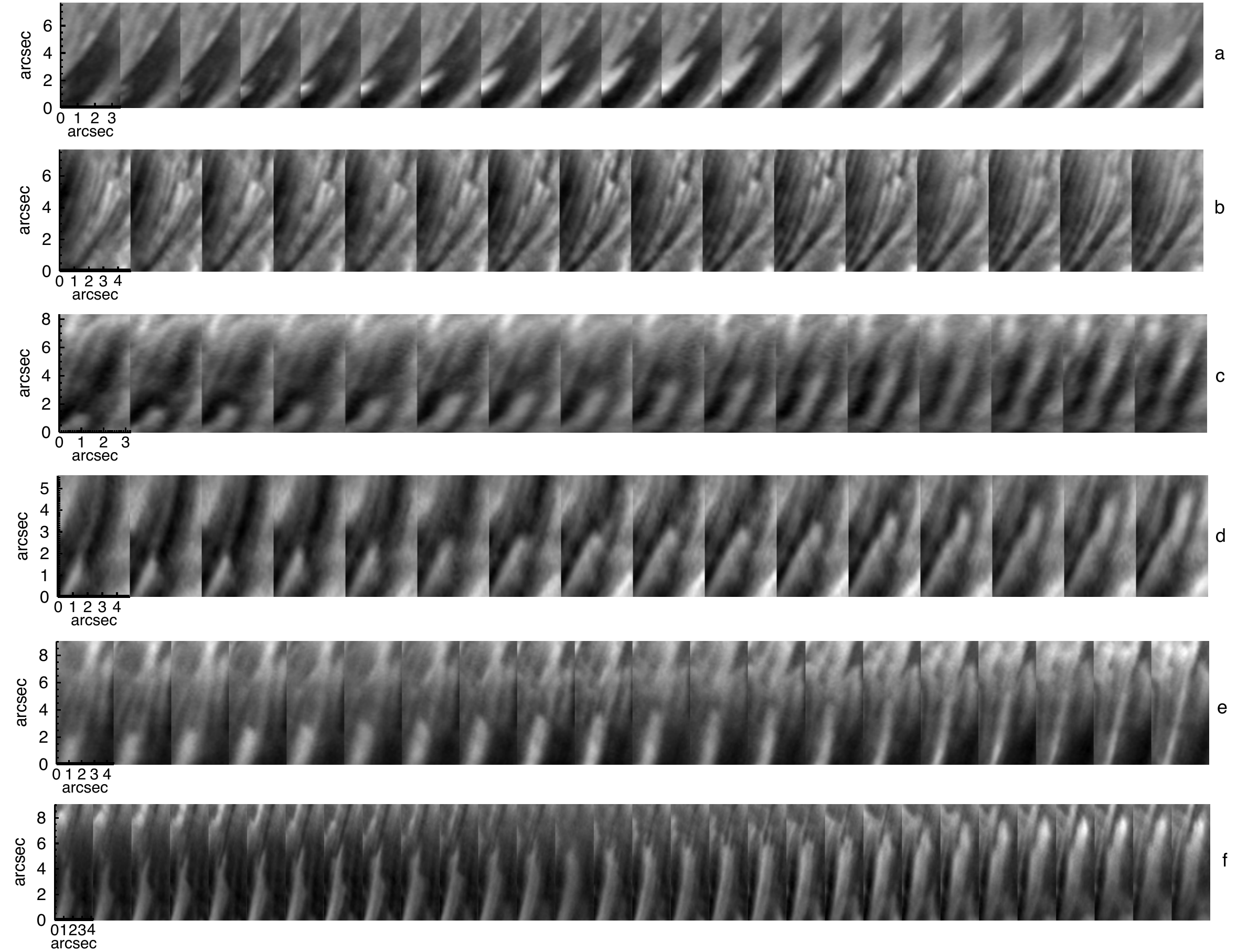}
\end{center}
\caption{Six examples of H$\alpha$ jets observed within the white rectangle in the 
lower-left panel of Fig~\ref{fig1}. Each successive image is separated by a 
time of 8.4~s, while the length of the jet structures range between approximately 
$3000$--$6000$~km.}
\label{fig2}
\end{figure*}

\section{Analysis and disscussion}
The area under investigation (see boxed region in the lower-left panel of Fig.~\ref{fig1}) 
consists of a multitude of highly dynamic 
thread-like structures, which connect areas of opposite magnetic polarity, forming a ``chromospheric arcade''.

\begin{figure}[t]
\begin{center}
\includegraphics[width=9cm]{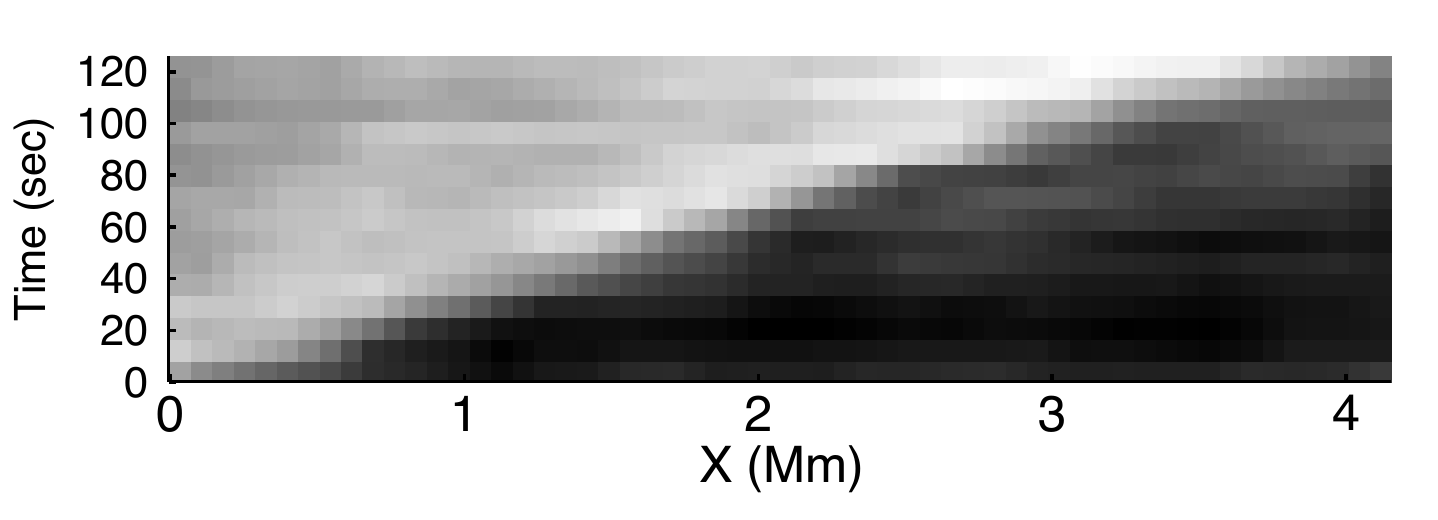}
\end{center}
\caption{H$\alpha$ time-distance diagram of jet $a$, taken from Fig.~\ref{fig2}. 
The gradient of the diagonal ridge in this $x$--$t$ slice corresponds to the jet velocity. The velocity of this jet is 34 km s$^{-1}$.}
\label{fig3}
\end{figure}

\begin{figure}[t]
\begin{center}
\includegraphics[width=8.4cm]{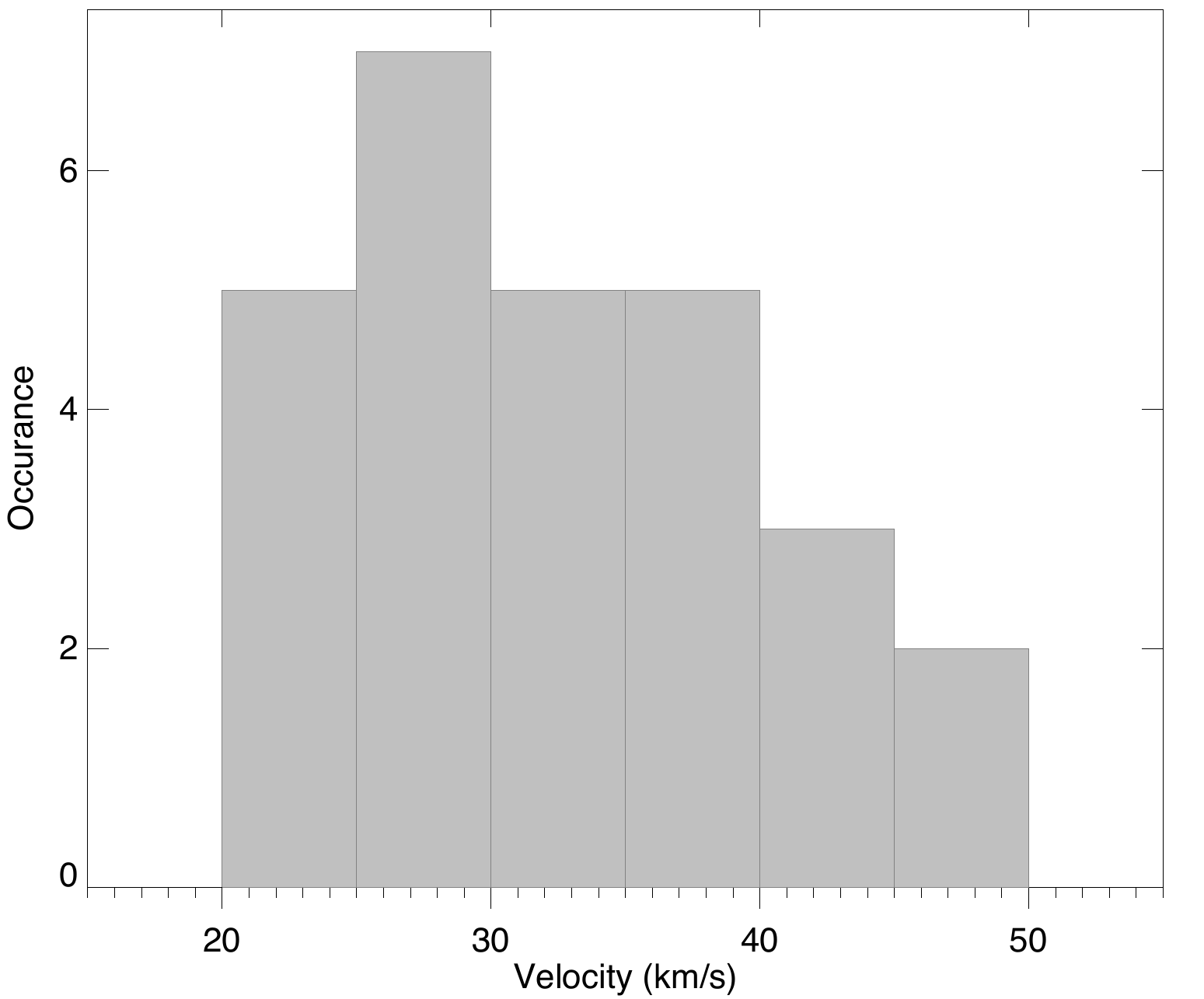}
\end{center}
\caption{Velocity distribution of the 27 H$\alpha$ jets.}
\label{fig4}
\end{figure}

A thorough examination of the H$\alpha$ core images reveals a total of 
27 jet-like structures, some of which are displayed in Fig.~\ref{fig2}. 
The velocities of the jets are determined using space-time slices 
($x$--$t$ plots), generated along the path of the jet trajectories 
(Fig.~\ref{fig3}). Velocity estimates for these jets are in the range 
$20$--$45$ kms$^{-1}$, with a histogram of their occurrence plotted 
in Fig.~\ref{fig4}. Detailed properties for the six most prominent jets 
(see Fig.~\ref{fig2}) are listed in Table~\ref{table1}.  

\begin{figure}[h]
\begin{center}
\includegraphics[width=8.6cm]{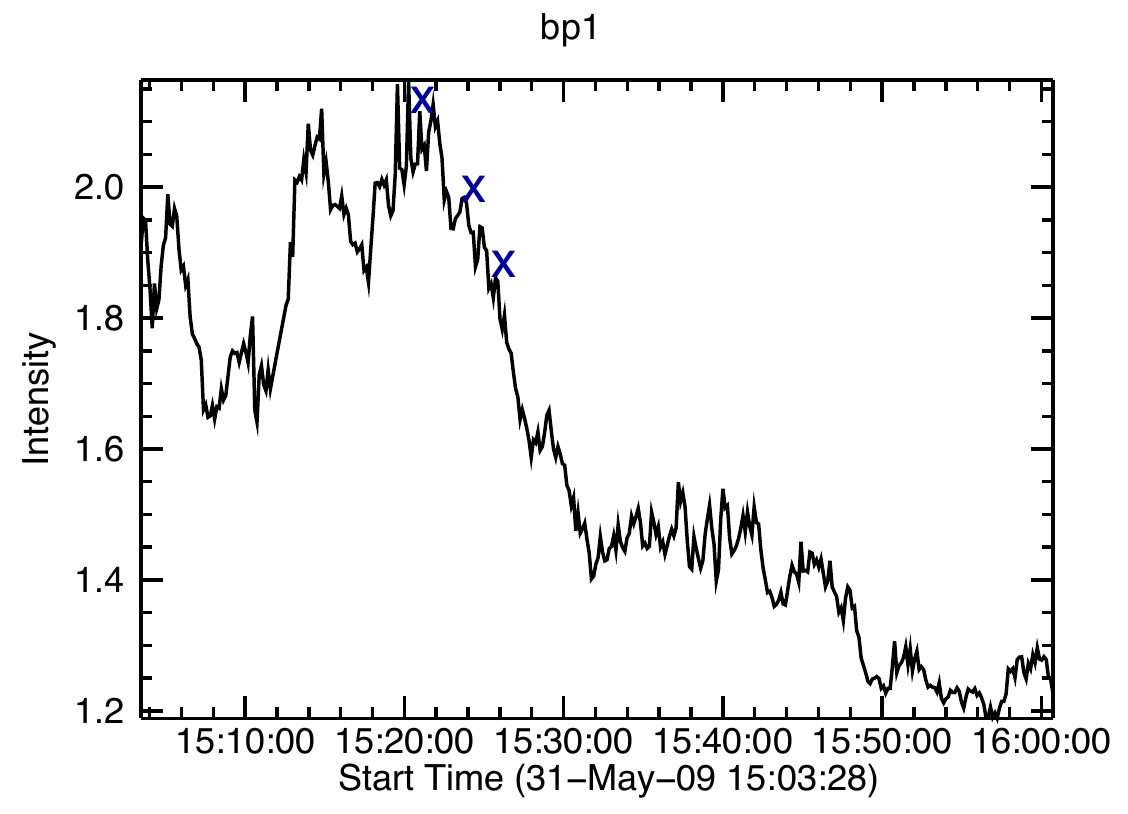}
\includegraphics[width=8.6cm]{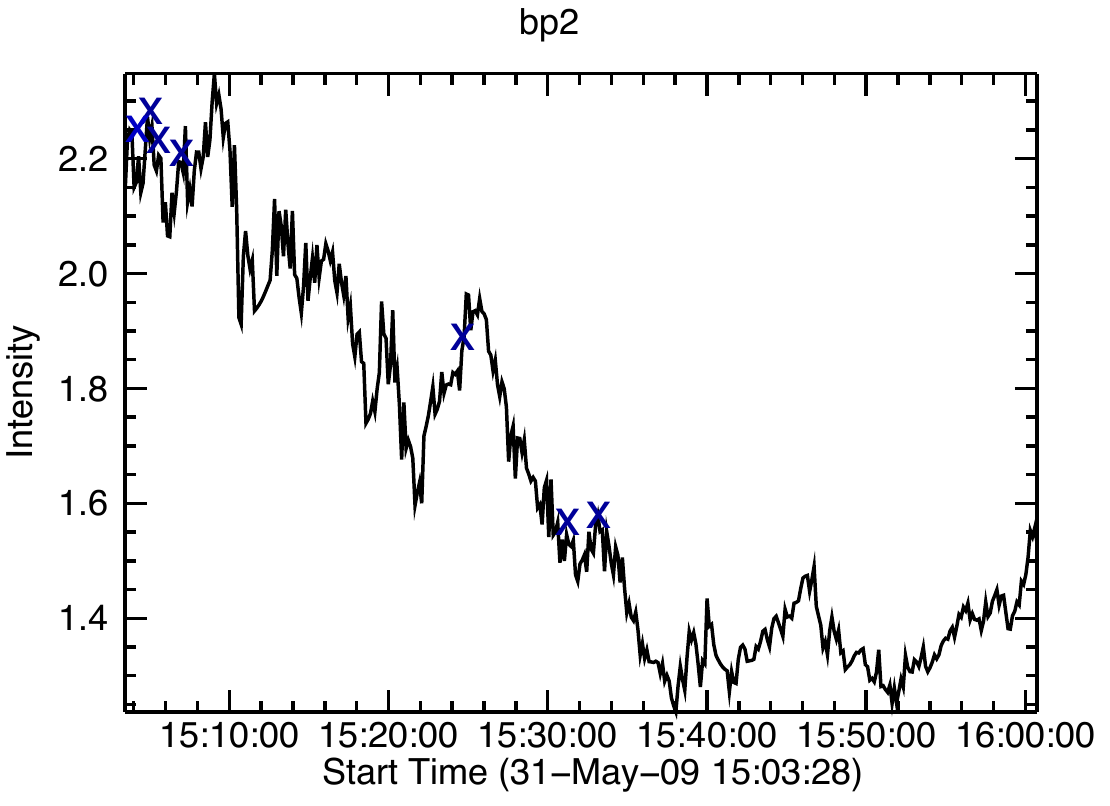}
\includegraphics[width=8.6cm]{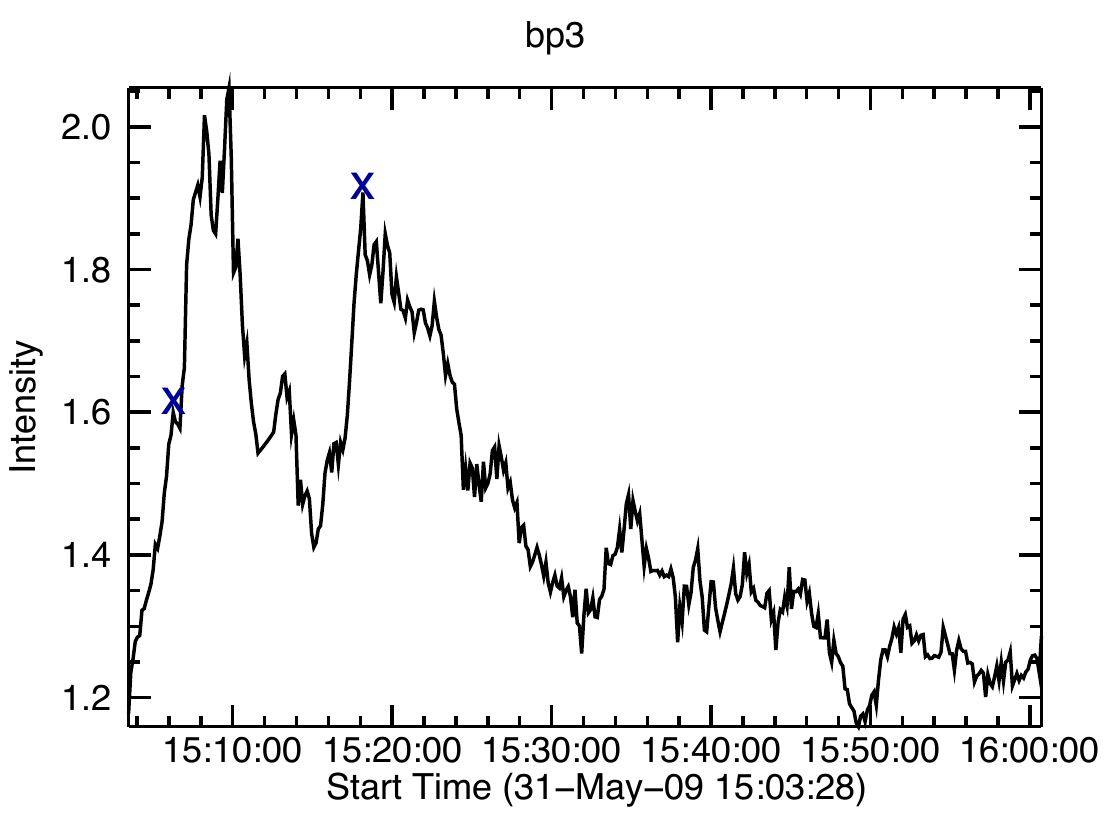}
\end{center}
\caption{Light curves of the Ca {\sc{ii}} K bright points outlined with rectangles in the 
lower-right panel of Fig.~\ref{fig1}.  `x' symbols indicate the time when a jet event originated from 
that particular bright point.}
\label{fig5}
\end{figure}

Almost half of the H$\alpha$ jets originate from locations co-spatial with Ca {\sc{ii}} K 
and G-band bright points.  Three examples of bright points are outlined in the lower-right 
panel of Fig.~\ref{fig1} using white rectangles and labelled as `bp1', `bp2', and 
`bp3'. All six jets shown in Fig.~\ref{fig2}, as well six additional jets, are co-spatial with these bright features, and  
the evolution of the Ca {\sc{ii}} K bright point intensities indicates a possible correlation with the H$\alpha$ jets.  
Ca {\sc{ii}} K intensity enhancements and strong isolated peaks are visible in the bright point light 
curves shown in Fig.~\ref{fig5}. 
In the light curve bp1 (top panel), the strongest peak around 15:20 UT 
is followed by three H$\alpha$ jets occurring at  $\sim$ 15:21, 15:24, 15:26 UT (crosses). From 15:29, there is no 
significant enhancement in the Ca II K intensity  and correspondingly no H$\alpha$ events.  
There is an enhanced intensity for the first few minutes in the light curve of bp2. 
During this time interval, four strong H$\alpha$ jets are generated at the location of bp2. 
Furthermore, one of the most isolated intensity peaks in the light curve of bp2 is around 15:25, 
and there is an H$\alpha$ jet observed very close to that time.  Two other  jets associated with the bp2 occur between 
15:31 and 15:33. A sharp decrease in the intensity of bp2 occurs after this time.  In the light curve of bp3, 
the peaks are located at around 15:10 and 15:18~UT. The eruption of two jets (e,f in the Fig.~\ref{fig2}) are observed near these peaks. 
H$\alpha$ jets appear to initiate at times when the Ca {\sc{ii}} K intensity rises, 
suggesting a possible link between the two phenomena. It must be noted that there are some bright peaks in the light curves of considered bright points (Fig.~\ref{fig5}) 
that are not accompanied by  H$\alpha$ jet events. In these cases the jets may not be visible due to overlapping H$\alpha$ structures. The remaining jets, of the 27 analysed in this work,  
also correspond to Ca {\sc{II}}  K brightenings, however, these brightenings are not observed as localised, simple, fine areas,and makes them somewhat more difficult to disentangle.

\begin{table}[t]
\caption{Detailed parameters of the jets displayed in Fig.~\ref{fig2}.}
\begin{center}
\begin{tabular}{c c c  c}
\hline
Event    &   Starting      &      Duration        &        Velocity   \\ 
name    &      time          &     of the event   &          (km/s)                 \\     
              &      (UT)        &          (sec) &                 \\  \hline

Jet a     &      15:24:46      &      160       &     34    \\ 
Jet b     &      15:03:53  &      142       &     20     \\
Jet c     &      15:31:36    &      135       &     35     \\
Jet d     &      15:33:08    &      145       &     38     \\
Jet e     &      15:18:20    &       177       &     21     \\
Jet f      &      15:05:27     &       250       &    23     \\
\hline
%\caption {ss}
\end{tabular}
\end{center}
\label{table1}
\end{table}%

Observed 27 jets are initiate  in about 9 distinct, small-scale ($\sim 5''\times5''$) regions.
 As  we have already mentioned, almost half of them originate from the three bright points labelled in the bottom right panel in Fig.~\ref{fig1}. 
This fact indicates repetitive nature of the observed jets, however, no well-defined period was found.

Light curves of background quiescent Ca {\sc{ii}} K regions 
do not indicate any intensity peaks or significant enhancements. We note that small amplitude 
intensity fluctuations consistent with the well-known 3-5 minute acoustic modes are observed in the both the 
quiescent and bright point regions (see e.g. Fleck \& Schmitz \cite{fleck}; Kariyappa et~al. \cite{kariyappa}). 
However, the intermittent nature of the jets and their supersonic velocities suggest that they are not associated with 
the acoustic mode oscillations. 

Striking similarities between H$\alpha$ core observations and 
TRACE Fe~{\sc{ix}}/{\sc{x}} images suggest that H$\alpha$ may be used as a proxy for coronal material. These similarities may be 
attributed to reduced H$\alpha$ opacity, excess emissivity, or a combination 
of both (Rutten~\cite{rutten}). A cool loop model may therefore 
be appropriate for chromospheric structures. 

Doyle et~al. (\cite{doyle}) suggested that one-directional plasma flow in a cold coronal loop  ($10^5$--$10^6$~K)  can be interpreted 
in terms of a short-lived siphon flow which may arise due to a  non-linear heating pulse at one of the loop footpoints. The supersonic flow velocities 
are determined by the duration of the pulse, in addition to its associated energy input. 
All our presented 
H$\alpha$ jets are one-directional, and propagate from negative (bottom portion of the white rectangle 
in the lower-left panel of Fig.~\ref{fig1}) to positive polarities (upper section of the white rectangle 
in the lower-left panel of Fig.~\ref{fig1}). The established supersonic velocities are very similar to those predicted by the siphon flow model.  
Furthermore,  other similarities between observations of Doyle et~al. (\cite{doyle}) and our findings, such as the comparable time scales 
of the observed events and also morphological similarities of the observed structures
motivated us to interpret our observation by means of siphon flow model.
We note that Uitenbroek et al. (\cite{han}) used spectroheliograms of the Ca II IR triplet (8542.1\AA) to detect a siphon flow associated with a magnetic pore.  Line of sight velocities as high as 27 km $\mathrm{s}^{-1}$ were detected.

Using the heating rate and plasma flow velocity given by the coronal loop simulation of Doyle et~al. (\cite{doyle}), we can estimate that the
rate of energy conversion from the source to the kinetic energy of the flow in the loop is $\sim 0.1$. This estimate is based on an 
ambient coronal density of $10^{-13}~\mathrm{g \cdot cm^{-3}}$. If we assume (i) that the same rate applies in the chromosphere, (ii) a chromospheric density 
of $10^{-11}~\mathrm{g \cdot cm^{-3}}$ and (iii) the chromospheric jet velocity of $40~\mathrm{km\,s^{-1}}$ as estimated from the observations, we obtain the total heating energy
of about $10~\mathrm{erg \cdot cm^{-3}}$. Assuming the duration of energy release in the chromospheric loop footpoint of the order of ten seconds,
we obtain a heating rate $h \sim 1 \mathrm{\,erg \cdot cm^{-3} s^{-1}}$,  a value significantly higher than the one estimated 
for the coronal loop. This is not surprising as more energy is required to propel the denser chromospheric jets to velocities comparable to those in the coronal model.
The  volume of the Ca {\sc{II}} K bright points (bottom right panel of the Fig.~\ref{fig1}), 
which are the footpoints of the H$\alpha$ jets, can be estimated as  $\mathrm{10^{24}\,  cm^{3}}$ 
(assuming a size of approximately  ~$2''\times 2''$  and a height of  $\sim 1'' $).   
Thus, the estimated released energy for the heating rate calculated above is about $10^{24}~\mathrm{erg}$. 
The total mass of the jet ($\sim10^{12}~\mathrm{g}$) is estimated from the chromospheric density and average volume of the jet. 
In order to bring the total mass up to a height of about $1.5\times 10^{3} \, \mathrm{km}$ an energy of $\sim \mathrm{10^{24}\, ~erg}$ is needed. 
These values are comparable to the estimated energy stored in the bright point volume.
We emphasize that the above estimates are approximate and a detailed simulation is required to account for all processes involved in this energy release.

\section{Concluding Remarks}%\label{sec:sum}
We present evidence for small-scale, one directional H$\alpha$ jets in the solar chromosphere. 
Jet velocities as high as 45~km s$^{-1}$ are found. 
Observation shows the relations between  H$\alpha$ jets and Ca {\sc{ii}} K bright points. 
We interpret this phenomena by means of a coronal loop siphon flow model, whereby 
one of the loop footpoints is driven by a localised, 
non-linear heating pulse (Doyle et~al. \cite{doyle}).
The  energy required to drive jets is estimated as  $10^{24}~\mathrm{erg}$.

The jets presented here appear to be associated 
with small-scale explosive events, and may provide a mechanism to support 
mass outflow from the chromosphere out into the corona.

\begin{acknowledgements}
Observations were obtained at the National Solar Observatory, operated by the 
Association of Universities for Research in Astronomy, Inc (AURA) under 
agreement with the National Science Foundation. This work is supported by the Science 
and Technology Facilities Council (STFC), with DBJ particularly grateful for the 
award of an STFC post-doctoral fellowship. We thank the Air Force Office of Scientific Research, Air Force 
Material Command, USAF for sponsorship under grant number FA8655-09-13085.
\end{acknowledgements}

\end{document}